# Room Temperature Quantum Spin Hall Insulators with a Buckled Square Lattice

Wei Luo[1] and Hongjun Xiang[1,2]*

[1]Key Laboratory of Computational Physical Sciences (Ministry of Education), State Key Laboratory of Surface Physics, and Department of Physics, Fudan University, Shanghai 200433, P. R. China

[2]Collaborative Innovation Center of Advanced Microstructures, Fudan University, Shanghai 200433, China

Email: hxiang@fudan.edu.cn

**ABSTRACT:** Two-dimensional (2D) topological insulators (TIs), also known as quantum spin Hall (QSH) insulators, are excellent candidates for coherent spin transport related applications because the edge states of 2D TIs are robust against nonmagnetic impurities since the only available backscattering channel is forbidden. Currently, most known 2D TIs are based on a hexagonal (specifically, honeycomb) lattice. Here, we propose that there exists the quantum spin Hall effect (QSHE) in a buckled square lattice. Through performing global structure optimization, we predict a new three-layer quasi-2D (Q2D) structure which has the lowest energy among all structures with the thickness less than 6.0 Å for the BiF system. It is identified to be a Q2D TI with a large band gap (0.69 eV). The electronic states of the Q2D BiF system near the Fermi level are mainly contributed by the middle Bi square lattice, which are sandwiched by two inert $BiF_2$ layers. This is beneficial since the interaction between a substrate and the Q2D material may not change the topological properties of the system, as we demonstrate in the case of the NaF substrate. Finally, we come up with a new tight-binding model for a two-orbital system with the buckled square lattice to explain the low-energy physics of the Q2D BiF material. Our study not only predicts a QSH insulator for realistic room temperature applications, but also provides a new lattice system for engineering topological states such as quantum anomalous Hall effect.

**Introduction**

Topological insulators (TIs),[1-7] as a new class of quantum materials, hold great potential for applications in quantum information, spintronics, field effect transistors, as well as thermoelectrics.[8-14] The most interesting character of TIs is the presence of robust topological Dirac surface states in three-dimensional (3D) TIs or helical edge states in two-dimensional (2D) TIs with spin locked to momentum as protected by time-reversal symmetry. For coherent nondissipative spin transport related applications, 2D TIs are better than 3D TIs as the electrons at the edges of 2D TIs can only move along two directions with opposite spins and thus free from backscattering caused by nonmagnetic defects, whereas the surface states of 3D TIs are only free from exact 180 °-backscattering and suffer from scattering of other angles.

Currently, quantum spin Hall effect (QSHE) was experimentally observed only at very low temperatures.[6,7] There is great interest in searching for room temperature (RT) 2D TIs. Among all predicted or synthesized 2D TIs, the hexagonal honeycomb lattice is the most common motif. For instance, graphene was predicted to be the first 2D topological insulator (TI).[15] However, QSHE in graphene may only be observed at an extremely low temperature due to its tiny band gap (about $8 \times 10^{-4}$ meV[16]) opened by spin-orbit coupling (SOC) effect. Subsequently, silicene with a buckled honeycomb lattice was predicted to be a 2D TI with a larger band gap (about 1.55 meV) due to its stronger intrinsic SOC effect and buckling effect.[10] Unfortunately, the band gap is still too small for RT application, and free-standing silicene or silicene grown on an insulating substrate have not been obtained experimentally. Wang *et al.* predicted that a 2D organometallic triphenyl-metal compound with the hexagonal lattice is a new organic TI with a band gap of about 8.6 meV.[14] Very recently, Song *et al.* proposed that 2D honeycomb Bi halide systems are huge-band-gap (up to 1.08 eV) TIs. These giant-gaps are entirely due to the result of strong on-site SOC being related to the $p_x$ and $p_y$ orbitals of the Bi atoms around the two valley K and K' of the honeycomb lattice.[17] Experiments are awaited to check the stability of 2D honeycomb Bi halides. We note that Wu *et al.* studied systematically the physics of the honeycomb lattice with $p_x$ and $p_y$ orbitals.[18-20] Although the presence of the Dirac band structure makes the honeycomb lattice the most common platform for engineering the topological states, a large gap 2D TI suitable for realistic RT applications remains to be synthesized experimentally. A nature question then arises: Does QSHE exist in other simple lattices? Addressing this question successfully will not only enrich our understanding of topological phenomena, but also pave new ways for designing topological materials for realistic applications.

In this work, we suggest that QSHE can occur in the buckled square lattice. With our newly developed global optimization approach, we predict that a new quasi-2D (Q2D) material (i.e., $Bi_4F_4$) with a distorted square lattice of Bi is much more stable than the previously proposed honeycomb BiF structure.[17] This $Bi_4F_4$ system is identified to be a room temperature TI with a large band gap (0.69 eV). A new tight-binding (TB) model for the buckled square lattice with two orbitals per site is proposed to understand the origin of the topological nontrivial electronic properties of $Bi_4F_4$.

**Results and discussions**

**Geometrical and electronic structures of the Q2D $Bi_4F_4$ material.** As we discussed above, the hypothetical 2D honeycomb BiF compound is found to be large-gap TIs[17]. However, it may not be the lowest energy structure. With our newly developed global optimization approach for predicting Q2D materials (see Methods),[21] we perform extensive search to find more stable Q2D BiF structures. Our calculations show that the Q2D BiF structure [shown in Fig. 1(a)] has the lowest energy among all structures with the thickness less than 6.0 Å. Since the unit cell contains four formula units of BiF, we refer to this structure as $Bi_4F_4$. Instead of the hexagonal lattice, $Bi_4F_4$ takes a distorted square lattice with two slightly different lateral lattice constants (4.520 Å and 4.527 Å). The total energy of this structure is lower than that of 2D honeycomb BiF[17] by about 0.42 eV/atom. The formation energy of $Bi_4F_4$ is calculated to be 1.62 eV/atom with respect to bulk Bi and fluoride molecule, indicating no phase separation behavior. Our Q2D $Bi_4F_4$ structure can be seen as a stacking of three layers: A bismuth layer with the distorted buckled square lattice is sandwiched between two $BiF_2$ layers. The computed phonon frequencies[22,23] (see Supplementary Fig. S1) indicate that $Bi_4F_4$ is dynamically stable. Furthermore, first-principles molecular dynamics simulations indicate that $Bi_4F_4$ is thermally stable at least up to 500 K (see Supplementary Fig. S2).

The local density approximation (LDA) band structure of $Bi_4F_4$ is shown in the left panel of Fig. 1(b). We can see a Dirac cone at the Fermi level close to the Y point. The detailed analysis shows that the bands around the Fermi level are mainly contributed by $p_x$ and $p_y$ orbitals of the middle Bi layer of $Bi_4F_4$. The $p_z$ orbitals of the middle Bi layer lie at around -1.5 eV and 2.0 eV with respect to the Fermi level (see Supplementary Fig. S3). We note that the electron-phonon coupling removes the Dirac cone near one X point and the other cone lying between $\Gamma$ and M point due to the presence of the in-plane distortion of the middle Bi layer. After including the SOC effect, the band gap opens at the Dirac point [see Fig. 1(b)]. The whole system has an indirect band gap of 0.37 eV with the conduction band minimum (CBM) near (0, 0.5). Since LDA usually underestimates the band gap, we adopt the more reliable hybrid functional HSE06[24] to find that the indirect band gap is 0.69 eV (see Supplementary Fig. S4). It is large enough for RT applications. Because $Bi_4F_4$ has an inversion symmetry, we can calculate the topological $Z_2$ invariant by counting the parity of the occupied states[25] and find that $Bi_4F_4$ is a 2D TI ($Z_2 = 1$).

For realistic applications, the Q2D $Bi_4F_4$ material should lie on a substrate, or $Bi_4F_4$ can be directly grown on a substrate. We find that the lattice constant of $Bi_4F_4$ is very close to that of the insulating substrate NaF. The lattice mismatch between NaF and $Bi_4F_4$ is about 2.0%. We suggest that the [001] NaF surface can be a good substrate for $Bi_4F_4$ [see Supplementary Fig. S5(a)]. Our band structure calculation shows that the whole system has a band gap and states near the Fermi level are mostly contributed by Bi $p_x$ and $p_y$ orbitals of $Bi_4F_4$ [see Supplementary Fig. S5(b)]. We find that $Bi_4F_4$ on the NaF substrate is still a TI. In order to confirm this, we gradually decrease the distance between $Bi_4F_4$ and the substrate NaF and find that the band gap does not close. In other words, the interaction between the insulating NaF substrate and $Bi_4F_4$ is weak so that the topological nontrivial electronic properties of $Bi_4F_4$ remain intact. It is noted that the two inert outer layers makes the topological properties of $Bi_4F_4$ more insensitive to the substrate, different from the cases of silicene and bi-layer Bi. We also check the robustness of the TI electronic structure of $Bi_4F_4$ with respect to an epitaxial strain. Our LDA+SOC calculations show that the indirect band gap of $Bi_4F_4$ increases with strain and there is a finite gap even when a 5% compressive strain is present (see Fig. 2). Note that the quasi-particle band gap should be larger since LDA underestimates the band gap.

**Electronic structures of a simple model system $Bi_2F_2$.** The main physics of $Bi_4F_4$ can be understood by considering a simpler system. As already mentioned above, $Bi_4F_4$ is composed by two $BiF_2$ layer and a middle Bi layer. The $BiF_2$ layer is inert since their electronic states are far from the Fermi level. The chemical bonding in the outer $BiF_2$ layer can be analyzed as follows. In a $BiF_2$ layer, each Bi atom is surrounded by four F atoms and forms a covalent bond with one Bi atom from the middle Bi layer; each F atom forms bonds with two Bi atoms of the $BiF_2$ layer. Among the five valence electrons ($6s^2 6p^3$) of a Bi atom, two electrons transfer to the neighboring F atoms, one electron is used to form a $p_z$-$p_z$ $\sigma$ bond with the Bi atom from the middle layer, and the remaining two electrons occupy the lone-pair 6s orbital. For a Bi atom in the middle Bi layer, one electron is used to form the $p_z$-$p_z$ $\sigma$ bond with Bi atom of $BiF_2$ layer, two electrons occupy the lone-pair 6s orbital, and the remaining two low-energy electrons will occupy the $p_x$ and $p_y$ orbitals. Therefore, we can adopt a simple model structure $Bi_2F_2$ (see Fig. 3) to understand the low-energy physics in $Bi_4F_4$. In this $Bi_2F_2$ structure, the Bi atoms form the buckled square lattices, neglecting the in-plane distortion. Each Bi atom is saturated with a fluorine atom to shift the $p_z$ orbital far from the Fermi level.

The LDA band structure of $Bi_2F_2$ is shown in the left panel of Fig. 3. Similarly to the case of $Bi_4F_4$, there are Dirac cones at the Fermi level. One difference is that there are more Dirac cones in $Bi_2F_2$ due to the absence of the in-plane distortion. Our analysis shows that the electronic states near the Fermi level are mostly contributed by Bi $p_x$ and $p_y$ orbitals. After including the SOC effect, the band gap is opened in $Bi_2F_2$, similar to the case of $Bi_4F_4$. Furthermore, the $Z_2$ invariant of $Bi_2F_2$ is also 1. Therefore, the simple $Bi_2F_2$ structure carries the main physical properties of the $Bi_4F_4$ structure, as we expected. To confirm further the TI behavior in Q2D BiF system with a square lattice, we carry out direct LDA+SOC calculations to obtain the edge band structure of a $Bi_2F_2$ nanoribbon with the width of 5.4 nm [see Supplement Fig. S6(a)]. We find there is a Dirac state with the linear dispersion mainly contributed by the edge Bi atoms at Γ point, evidencing the topological nontrivial behavior [see Supplement Fig. S6(b)].

**TB model for the buckled square lattices.** In order to understand the topological electronic structures of $Bi_2F_2$ and $Bi_4F_4$, we propose a new TB model for the buckled square lattice, as shown in Fig. 4(a). Because of the presence of buckling, there are two lattice points in the primitive cell and the lattice is divided into two sublattices [A (top) and B (bottom)]. This square lattice has an in-plane lattice spacing d and, for simplicity, we assume d = 1 in the following discussion. We can see that the lateral mirror symmetry is broken due to the buckling, but the system still has spatial inversion symmetry at the bond centers between the nearest-neighbor (NN) A and B sites. At each lattice point, there are two orbitals $p_x$ and $p_y$. The number of electrons in a primitive cell is four, i.e., half-filled case. The total effective low-energy Hamiltonian contains three parts: $H = H_t + H_{so} + H_1$, with $H_t = \sum_{i \in A,s} [t_{pp\sigma}(c^+_{i,xs}c_{i\pm\hat{x},xs} + c^+_{i,ys}c_{i\pm\hat{y},ys}) + t_{pp\pi}(c^+_{i,xs}c_{i\pm\hat{y},xs} + c^+_{i,ys}c_{i\pm\hat{x},ys})] + H.c.$,

$H_{so} = \sum_{i \in \{A,B\}} (-\frac{i\lambda}{2}c^+_{i,x\uparrow}c_{i,y\uparrow} + \frac{i\lambda}{2}c^+_{i,x\downarrow}c_{i,y\downarrow}) + H.c.$, and $H_1 = \sum_{i \in A,\hat{d},\alpha,s} t^{eff}_{\hat{d},\alpha s} c^+_{i,\alpha s}c_{i+\hat{d},\alpha\bar{s}} + H.c.$,

where $c_{i,\alpha s}$ destructs a particle with spin s (↑ or ↓ along the z axis) on the $\alpha$ ($p_x$ or $p_y$) orbital at the site i; $\hat{x} \equiv (1,0)$ and $\hat{y} \equiv (0,1)$ represent the unit vectors in the x and y directions, respectively. The first term $H_t$ represents the usual NN hopping term between the p orbitals with the Slater-Koster interaction parameters $t_{pp\sigma}$ (>0) and $t_{pp\pi}$ (<0).[26] The second term is the on-site SOC ($\lambda \vec{L} \cdot \vec{S}$ with $\lambda$ as the magnitude of SOC effect) between the $p_x$ and $p_y$ orbitals written in a second quantization form.

The third term $H_1$ is a new additional effective NN hopping caused by the SOC effect, where $\bar{s}$ represents the spin state that is opposite to $s$, $\bar{\alpha}$ labels the orbital different from $\alpha$, and $\hat{d}$ is the in-plane distance vector between two NN lattice points. The strength of the effective hopping is described by sixteen interaction parameters $t^{eff}_{\hat{d},\alpha s}$. These parameters are related to each other by symmetry: $t^{eff}_{\hat{x},x\uparrow} = -it^{eff}$, $t^{eff}_{\hat{y},x\uparrow} = -t^{eff}$, $t^{eff}_{\hat{d},\alpha s} = -(t^{eff}_{\hat{d},\alpha \bar{s}})^*$, $t^{eff}_{\hat{d},\alpha s} = -t^{eff}_{-\hat{d},\alpha s}$, $t^{eff}_{\hat{d},\alpha s} = (t^{eff}_{\hat{d},\bar{\alpha} s})^*$, where $t^{eff}$ (>0) represents the magnitude of the effective hopping. Physically, this effective hopping may originate from a two-step process. As an example, we show how this works for the case of $t^{eff}_{\hat{x},y\uparrow}$ in Fig. 4(c). Although the basis of our low-energy Hamiltonian only contains $p_x$ and $p_y$ orbitals, there are high-lying $p_z$ orbitals in reality. In the first step, a spin-up electron on the $p_y$ orbital of the A sublattice hops into the spin-down $p_z$ orbital due to the on-site SOC. Then, this electron transfers to the spin-down $p_y$ orbital of the NN B-site, which is possible due to the presence of the buckling. Note that the second step is forbidden in a planar square lattice, which makes the effective NN hopping term vanishing there. Overall, the effective NN hopping term is a first-order SOC effect with $t^{eff} \propto \lambda$. This two-step process is different from the cases of systems with the honeycomb lattice. In graphene and silicene, the relevant orbital is $p_z$ and there is effective next-NN hopping due to the second-order[16] or first-order SOC[10] effect, respectively. On the other hand, the on-site SOC was found to be responsible for the band gap opening in honeycomb Bi halides with the low-energy $p_x$ and $p_y$ orbitals.[17] In the following, we will show that the effective NN hopping $H_1$ plays a key role in realizing the TI state.

We now discuss the electronic structure of our newly proposed model. First, we only consider the usual hopping term $H_t$. A typical band structure with $t_{pp\sigma} = 1.0$ and $t_{pp\pi} = 0.2$ is shown in the left panel of Fig. 4(b). We can see that there are two kinds of Dirac cones at the Fermi level; one locates at X point, another lies between Γ and M. These Dirac cones can be seen as a result of the Brillouin zone folding since $H_t$ does not break the translation symmetry along $\hat{x}$ and $\hat{y}$. We find that all four degenerate occupied bands at Γ have even parity, while there are two occupied even-parity bands and two occupied odd-parity bands for the other three time-reversal-invariant k-points (two X and one M). The fact that the states with even parity have lower energy than those with odd parity at Γ can be easily understood by the orbital interaction picture (see Supplementary Fig. S7). Second, we add the on-site SOC term $H_{so}$. The band structure in the case of $\lambda = 0.6$ is shown in the middle panel of Fig. 4(b). Obviously, the Dirac cones are kept although their positions are shifted. Therefore, the on-site SOC effect can not open a band gap in the buckled square lattices, in contrast to the case of honeycomb Bi halides.[17] Some bands split due to the on-site SOC interaction. In particular, the four-fold degenerate occupied bands at Γ split into two two-fold degenerate states whose parities remain even.

Finally, we find that the effective NN hopping $H_1$ can open a band gap at the Dirac points, as shown in the right panel of Fig. 4(b) for the case of $t^{eff} = 0.2$. The magnitude of the band gap is almost proportional to $t^{eff}$ when fixing other parameters [see Fig. 4(e)]. Note that there is no gap if we switch off the on-site SOC term. An important finding is that all four occupied states at Γ have even parity, while the number of occupied states with even-parity is the same as that with odd-parity for X and M. According to the parity criteria,[25] the system is a 2D TI ($Z_2 = 1$). Therefore, we find that the effective NN hopping due to SOC is indispensable for the occurrence of the QSHE.

We now investigate the dependence of the ground state on the parameters of the model Hamiltonian in details. As we discussed above, the system is not a TI whenever $t^{eff} = 0$. Now, we consider the case of $t^{eff} > 0$. We find that the system becomes a trivial insulator when the on-site SOC is too strong. This can be understood as follows. Since the parities of the occupied states at X and M do not depend on $\lambda$, we can focus on the eigenstates at Γ. By solving exactly the full model Hamiltonian at Γ, we obtain four two-fold degenerate states: $E_I = -\lambda/2 - 2(t_{pp\sigma} + t_{pp\pi})$, $E_{II} = \lambda/2 - 2(t_{pp\sigma} + t_{pp\pi})$, $E_{III} = -\lambda/2 + 2(t_{pp\sigma} + t_{pp\pi})$, $E_{IV} = \lambda/2 + 2(t_{pp\sigma} + t_{pp\pi})$. States I and II have even parity, while states III and IV have odd parity. Interestingly, these eigenvalues are independent on $t^{eff}$. Since $\lambda > 0$ and usually $t_{pp\sigma} + t_{pp\pi} > 0$, state I with the lowest energy is always occupied, while state IV with the highest energy is always empty. Whether state II has a lower energy than state III depends on the relative magnitude of sign of $\lambda/4$ and $t_{pp\sigma} + t_{pp\pi}$. When the on-site SOC is weak, i.e., $\lambda/4 < t_{pp\sigma} + t_{pp\pi}$, state II with even parity is occupied and the system is in a TI state. On the contrary, state III with odd parity is occupied and the system is a trivial insulator. The phase diagram in the case of $t^{eff} > 0$ is shown in Fig. 4(d). Noted that our model is different from that of FeSe system proposed by Hao *et al.*[27] Their model is intended to describe the interactions between d orbitals, the band gap induced by the SOC is small, and the appearance of the topological phase needs the help of a substrate.

We find that the band structure of $Bi_2F_2$ is very similar to that of our new TB Hamiltonian [compare Fig. 4(b) and Fig. 3]. It is anticipated that the low-energy physics of $Bi_2F_2$ can be described by our model. In order to prove this point, we obtain the low-energy effective Hamiltonian by using the maximally-localized Wannier function (MLWF).[28-30] We choose $p_x$ and $p_y$ orbitals as the projection functions for each Bi atom. The band structure near the Fermi level is reproduced rather well by the MLWF Hamiltonian (see the right panel of Fig. 3, denoted by blue lines). With the MLWF basis, we can obtain the parameters of our above TB model Hamiltonian: $t_{pp\sigma} = 1.34$ eV, $t_{pp\pi} = -0.26$ eV, $\lambda = 1.40$ eV, $t^{eff} = 0.13$ eV. Other interactions such as further neighboring hopping are found to be much weaker. Therefore, $Bi_2F_2$ and thus $Bi_4F_4$ can be well described by our model for the buckled square lattice.

**Conclusions.** Using the new particle swarm optimization (PSO) approach adapted for the Q2D systems, we predict that the most stable Q2D structure with the thickness less than 6 Å is a three-layer structure (i.e., $Bi_4F_4$) for BiF, in which the middle Bi layer takes a distorted square lattice. We find that $Bi_4F_4$ is a RT TI with a band gap of 0.69 eV. The TI state in $Bi_4F_4$ is robust with respect to the interaction with substrate and epitaxial strain. In order to understand its topological electronic structures, we proposed a TB model for the buckled square lattice with two orbitals (i.e., $p_x$ and $p_y$) per site. We show that the band inversion is due to the chemical hybridization between the $p_x$ and $p_y$ orbitals. Different from the cases of 2D systems (graphene, silicene, honeycomb Bi halides) with a honeycomb lattice, a newly discovered effective NN hopping term due to the SOC is crucial for the gap opening of Dirac cone. Bonding analysis and MLWF calculations suggest that the low-energy electronic structure of $Bi_4F_4$ can be described rather well by the new TB model. In this work, we not only predict a RT quantum spin Hall insulator with a distorted square lattice for realistic spintronic applications, but also provide a new basic TB model for investigating topological states in the simple square lattice.

**Methods.**

**DFT calculations.** In our work, density functional theory (DFT) method is used for structural relaxation and electronic structure calculation. The ion-electron interaction is treated by the projector augmented-wave[31] technique as implemented in the Vienna ab initio simulation package.[32] The exchange-correlation potential is treated by LDA.[33,34] For structural relaxation, all the atoms are allowed to relax until atomic forces are smaller than 0.01 eV/Å. The HSE06 functional is adopted to

compute the accurate band gap.[24]

**PSO Algorithm for Q2D Systems.** To obtain the most stable Q2D structure of BiF, we adopt our newly developed global optimization PSO approach for predicting Q2D materials.[21] In our implementation, we generate random Q2D structures with the thickness along the c-axis smaller than a given thickness. For each Q2D structure, we first randomly select a layer group instead of a planar space groups.[35] The lateral lattice parameters and atomic positions are then randomly generated but confined within the chosen layer group symmetry. Subsequently, local optimization including the atomic coordinates and lateral lattice parameters is performed for each of the initial structures. In the next generation, a certain number of new structures (the best 60% of the population size) are generated by PSO.[36] The other structures are generated randomly, which is critical to increase the structure diversity. When we obtain the new Q2D structure by the PSO operation or random generation, we make sure that the thickness of the Q2D structure is smaller than the given thickness.

In our simulation for the BiF system, we set the population size to 30 and the number of generations to 20. The ratio of bismuth and fluoride is fixed to 1:1 and the chemical formula ranges from $Bi_1F_1$ to $Bi_8F_8$. We consider six different initial thickness (between 1 and 6 Å) for each system. In addition, we repeat twice of each calculation in order to make results reliable.

## ASSOCIATED CONTENT

**Supplementary Information**

Supplementary Information accompanies this paper at http://pubs.acs.org.

Correspondence and requests for materials should be addressed to H.X.

## AUTHOR INFORMATION
Corresponding Author
*E-mail: hxiang@fudan.edu.cn (H. J. Xiang).

## ACKNOWLEDGMENTS

Work was supported by NSFC, the Special Funds for Major State Basic Research, FANEDD, NCET-10-0351, Research Program of Shanghai Municipality and MOE, Program for Professor of Special Appointment (Eastern Scholar), and Fok Ying Tung Education Foundation.

**REFERENCES**

1. Hasan, M. Z.; Kane, C. L. Rev. Mod. Phys. **2010,** 82, 3045.
2. Qi, X. L.; Zhang, S. C. Rev. Mod. Phys. **2011,** 83, 1057.
3. Yan, B. H.; Zhang, S.-C. Rep. Prog. Phys. **2012,** 75, 096501.
4. Qian, X.; Liu, J.; Fu, L.; Li, J. Science **2014,** 346, 1344.
5. Bernevig, B. A.; Hughes, T. L.; Zhang, S. C. Science **2006,** 314, 1757.
6. König, M.; Wiedmann, S.; Brüne, C.; Roth, A.; Buhmann, H.; Molenkamp, L. W.; Qi, X. L.; Zhang, S. C. Science **2007,** 318, 766.
7. Knez, I.; Du, R. R.; Sullivan, G. Phys. Rev. Lett. **2011,** 107, 136603
8. Weng, H. M.; Dai, X.; Fang, Z. Phys. Rev. X **2014,** 4, 011002.
9. Kane, C. L.; Mele, E. J. Phys. Rev. Lett. **2005,** 95, 146802.
10. Liu, C. C.; Feng, W. X.; Yao, Y. G. Phys. Rev. Lett. **2011,** 107, 076802.
11. Murakami, S. Phys. Rev. Lett. **2006,** 97, 236805.
12. Chuang, F. C.; Yao, L. Z.; Huang, Z. Q.; Liu, Y. T.; Hsu, C. H.; Das, T.; Lin, H.; Bansil, A. Nano Lett. **2014,** 14, 2505.
13. Xu, Y.; Yan, B. H.; Zhang, H. J.; Wang, J.; Xu, G.; Tang, P. Z.; Duan, W. H.; Zhang, S. C. Phys. Rev. Lett. **2013,** 111, 136804.
14. Wang, Z. F.; Liu, Z.; Liu, F. Nat. Commun. **2012,** 4, 1471.
15. Kane, C. L.; Mele, E. J. Phys. Rev. Lett. **2005,** 95, 226801.
16. Yao, Y. G.; Ye, F.; Qi, X. L.; Zhang, S. C.; Fang, Z. Phys. Rev. B. **2007,** 75, 041401.
17. Song, Z. G.; Liu, C. C.; Yang, J. B.; Han, J. Z.; Fu, B. T.; Ye, M.; Yang, Y. C.; Niu, Q.; Lu, J.; Yao, Y. G. NPG Asia Materials, **2014,** 6, e147.
18. Wu, C. J.; Bergman, D.; Balents, L.; Sarma, S. D. Phys. Rev. Lett. **2007**, 99, 070401.

19. Wu, C. J. Phys. Rev. Lett. **2008**, 101, 186807.

20. Zhang, G. F.; Li, Y.; Wu, C. J. Phys. Rev. B. **2014**, 90, 075114.

21. Luo, W.; Ma, Y. M.; Gong, X. G.; Xiang, H. J. J. Am. Chem. Soc. **2014,** 136, 15992.

22. Parlinski, K.; Li, Z. Q.; Kawazoe, Y. Phys. Rev. Lett. **1997,** 78, 4063.

23. Togo, A.; Oba, F.; Tanaka, I. Phys. Rev. B **2008,** 78, 134106.

24. Heyd, J. Scuseria, J. E. J. Chem. Phys. **2003,** 118, 8207.

25. Fu, L.; Kane, C. L. Phys. Rev. B. **2007,** 76, 045302.

26. Slater, J. C.; Koster, G. F. Phys. Rev. **1954,** 94, 1498.

27. Hao, N. N.; Hu, J. P. Phys. Rev. X **2014,** 4, 031053.

28. Mostofi, A. A.; Jonathan, R. Y.; Lee, Y. S.; Souza, I.; Vanderbilt, D.; Marzari, N. Comput. Phys. Commun. **2008,** 178, 685.

29. Marzari, N.; Vanderbilt, D. Phys. Rev. B **1997,** 56, 12847.

30. Souza, I.; Marzari, N.; Vanderbilt, D. Phys. Rev. B **2001,** 65, 035109.

31. Blöchl, P. E. Phys. Rev. B **1994,** 50, 17953.

32. Kresse, G.; Hafner, J. Phys. Rev. B **1994,** 49, 14251.

33. Perdew, J. P.; Zunger, A. Phys. Rev. B **1981,** 23, 5048.

34. Ceperkey, D. M.; Alder, B. J. Phys. Rev. Lett. **1980,** 45, 566.

35. Luo, X. Y.; Yang, J. H.; Liu, H. Y.; Wu, X. J.; Wang, Y. C.; Ma, Y. M.; Wei, S. H.; Gong, X. G.; Xiang, H. J. J. Am. Chem. Soc. **2011,** 133, 16285.

36. Wang, Y. C.; Lv, J.; Zhu, L.; Ma, Y. M. Phys. Rev. B **2010,** 82, 094116.

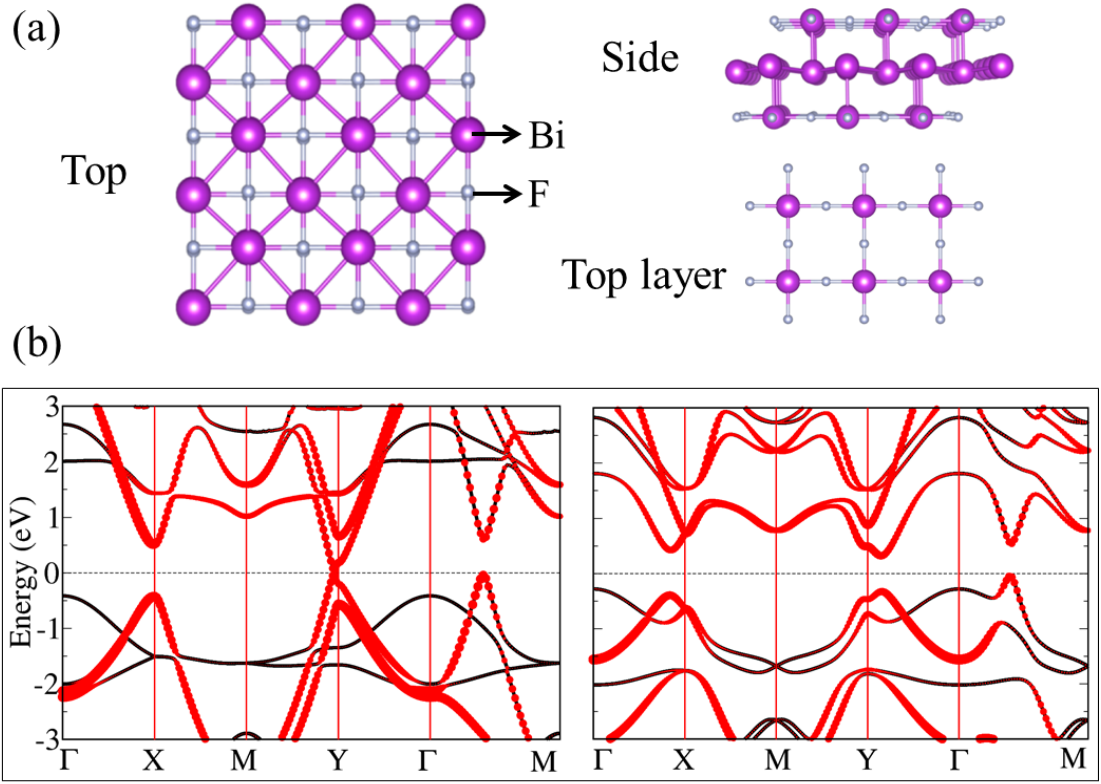

**Figure 1.** Geometrical and electronic structures of $Bi_4F_4$. (a) Top and side view of $Bi_4F_4$. The top and bottom layers are inert $BiF_2$ layers. The square lattice of the middle Bi layer is slightly distorted. (b) LDA band structures of $Bi_4F_4$ without (left panel) and with (right panel) SOC. The contribution from $p_x$ and $p_y$ orbitals of the middle Bi layer is highlighted.

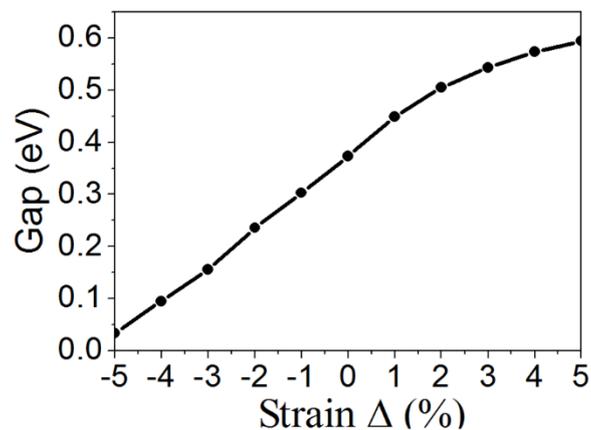

**Figure 2.** Band gap of $Bi_4F_4$ as a function of epitaxial strain from the LDA+SOC calculations. The indirect band gap increases with strain $\Delta$ ($\Delta = (a - a_0)/a_0$, where

$a$ and $a_0$ is the lateral lattice constants of the strained and unstrained systems, respectively). Even a large compressive strain (about -5%) can not destroy the nontrivial topological properties, indicating the robust TI behavior in $Bi_4F_4$. Note that LDA underestimates the band gap.

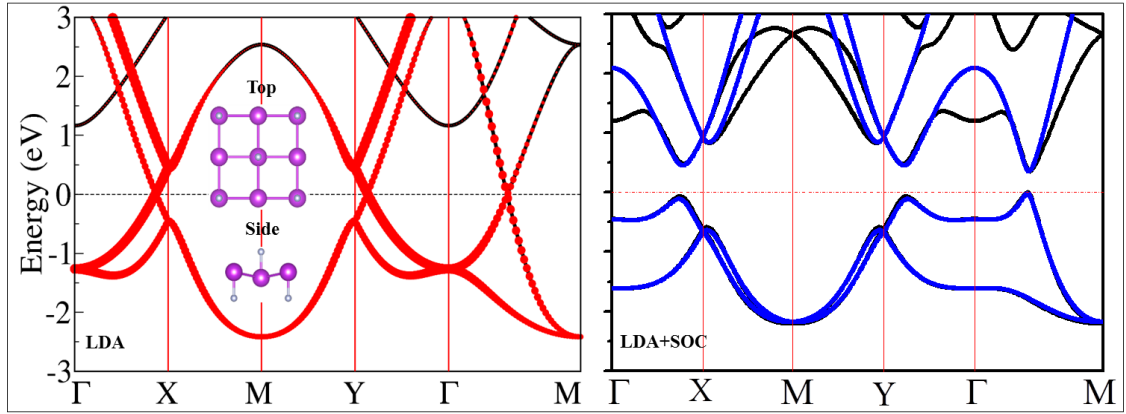

**Figure 3.** Geometrical and electronic structures of $Bi_2F_2$ with a buckled square lattice. Left panel: LDA band structure without SOC. There are more Dirac cones compared with Fig. 1(b) since the square lattice in $Bi_2F_2$ is not distorted. The contribution from $p_x$ and $p_y$ orbitals of Bi is highlighted. The inset shows the top and side views of $Bi_2F_2$. Right panel: Band structure from the LDA+SOC calculations. The band structure obtained by the MLWF Hamiltonian is shown in blue.

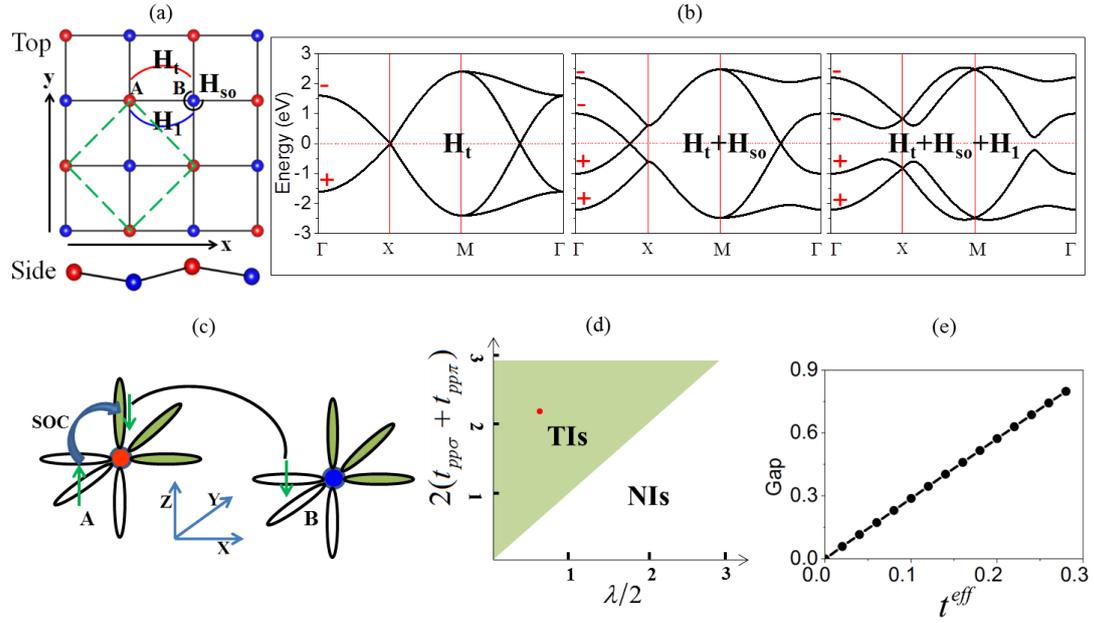

**Figure 4.** Description and electronic properties of our new TB model for a buckled square lattice. (a) Schematic illustration of the buckled square lattice. The red and blue balls represent sublattices A (top) and B (bottom), respectively. One space inversion center in the middle of NN A and B sites is denoted by "i". (b) Band structures of the buckled square lattice in the cases of the usual hopping ($H_t$, left panel), usual hopping and on-site SOC ($H_t+H_{so}$, middle panel), and all three terms ($H_t+H_{so}+H_1$, the right panel), respectively. The parities of states at $\Gamma$ are marked. (c) Illustration of how a first order SOC effect induces the effective NN hopping for the case of $t^{eff}_{\hat{x},y\uparrow}$. (d) The phase diagram of the TB model for a positive $t^{eff}$. The olive region represents TI. The red dot denotes the case of $Bi_2F_2$. (e) Band gap as a function of the effective NN hopping $t^{eff}$ when fixing other parameters.

*Supplementary Information for*

"Room Temperature Quantum Spin Hall Insulators with a Buckled Square Lattice"

Wei Luo[1] and Hongjun Xiang[1,2]*

[1]Key Laboratory of Computational Physical Sciences (Ministry of Education), State Key Laboratory of Surface Physics, and Department of Physics, Fudan University, Shanghai 200433, P. R. China

[2]Collaborative Innovation Center of Advanced Microstructures, Fudan University, Shanghai 200433, China

[*]Email: hxiang@fudan.edu.cn

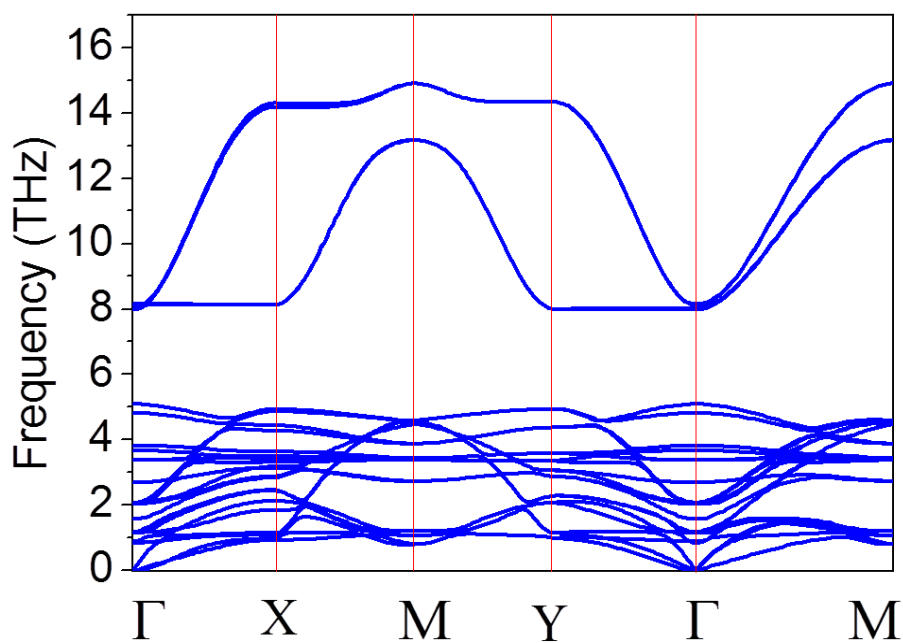

**Supplementary Figure S1. The phonon dispersions of $Bi_4F_4$ from the LDA calculations**. There are no imaginary frequencies in phonon dispersion, indicating its dynamical stability.

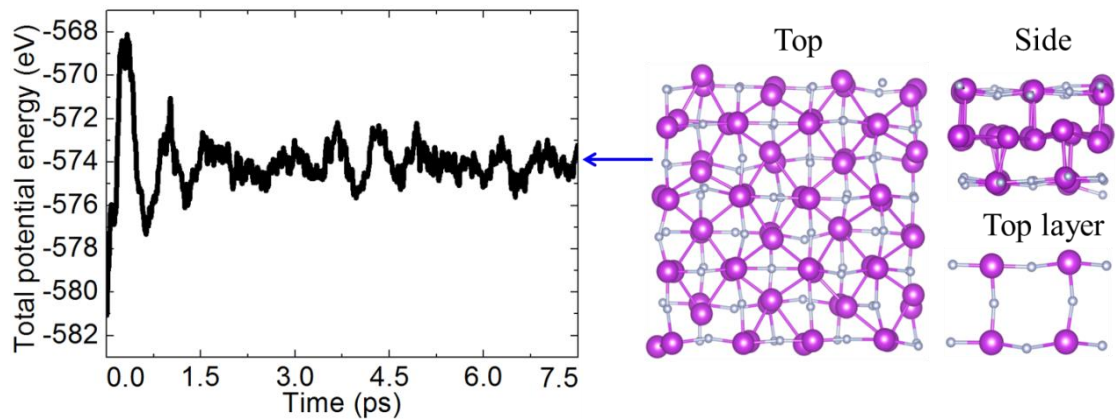

**Supplementary Figure S2. The thermal stability of $Bi_4F_4$.** The left figure is the fluctuation of total energy (128-atoms/cell) as a function of simulation time from first-principles canonical molecular dynamics simulations at 500 K. The right figure shows the final structure after the molecular dynamics simulation. One can see that the basic framework is kept intact, indicating that $Bi_4F_4$ is thermally stable at 500K.

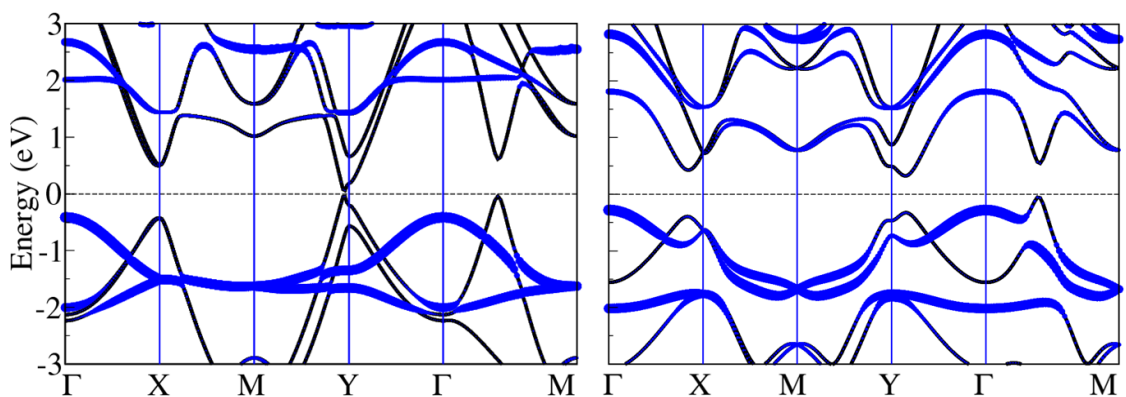

**Supplementary Figure S3. Band structures of $Bi_4F_4$ obtained from the LDA (left panel) and LDA+SOC (right panel) calculations, respectively.** The contribution from $p_z$ orbitals of the middle Bi layer is highlighted.

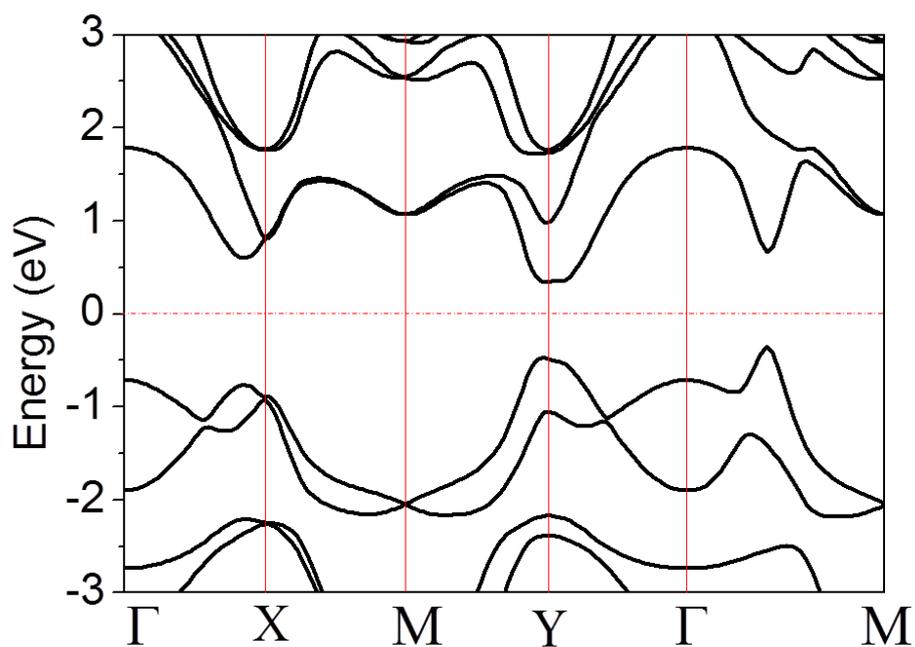

**Supplementary Figure S4. The band structure of $Bi_4F_4$ calculated by HSE06+SOC.** One can see that the band structure is similar to that in the case of LDA+SOC. The accurate band gap of $Bi_4F_4$ is about 0.69 eV from the HSE06+SOC calculation. This is large enough for RT application. Note that LDA+SOC gives a smaller band gap since LDA usually underestimates band gaps.

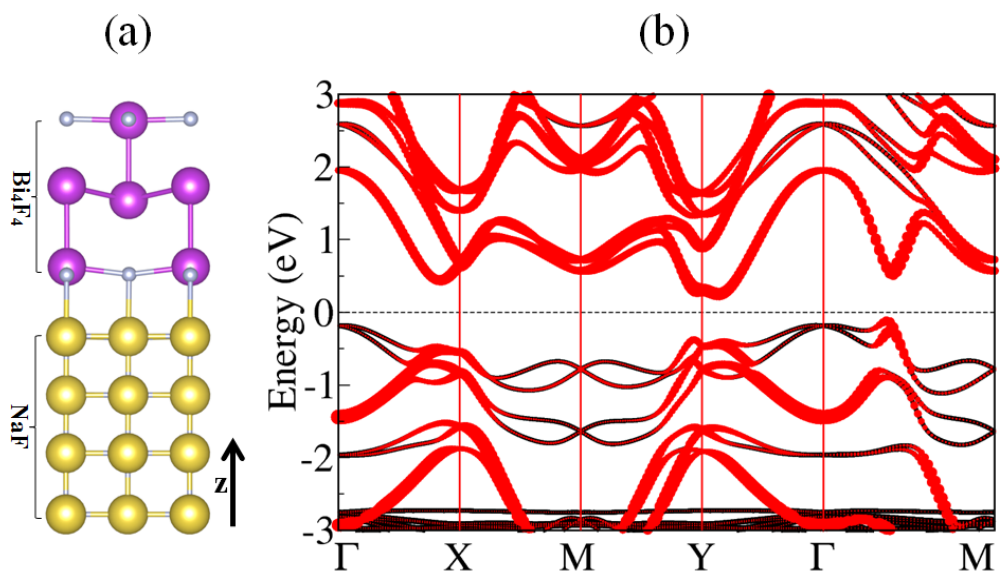

**Supplementary Figure S5. Geometrical and electronic structures of $Bi_4F_4$ on NaF**

**substrate.** (a) The structure of $Bi_4F_4$ on the [001] NaF surface after structural optimization. NaF takes a rock salt structure. The Bi (F) atom of $Bi_4F_4$ is on top of the F (Na) atom of the NaF substrate due to the electrostatic interaction. The distance between $Bi_4F_4$ and the surface is about 2.62 Å. (b) Band structure of the whole system calculated by LDA+SOC. The states near the Fermi level is mostly contributed by $p_x$ and $p_y$ orbitals (denoted by red circles) of the middle Bi layer of $Bi_4F_4$. This is because NaF is a large band gap (about 5.90 eV from LDA) insulator with a very low valence band and a very high conduction band. The band structure of the whole system near the Fermi level is similar to that of the free-standing $Bi_4F_4$ except for some band splitting due to the extrinsic Rashba effect. This suggests that $Bi_4F_4$ on NaF substrate is still a TI.

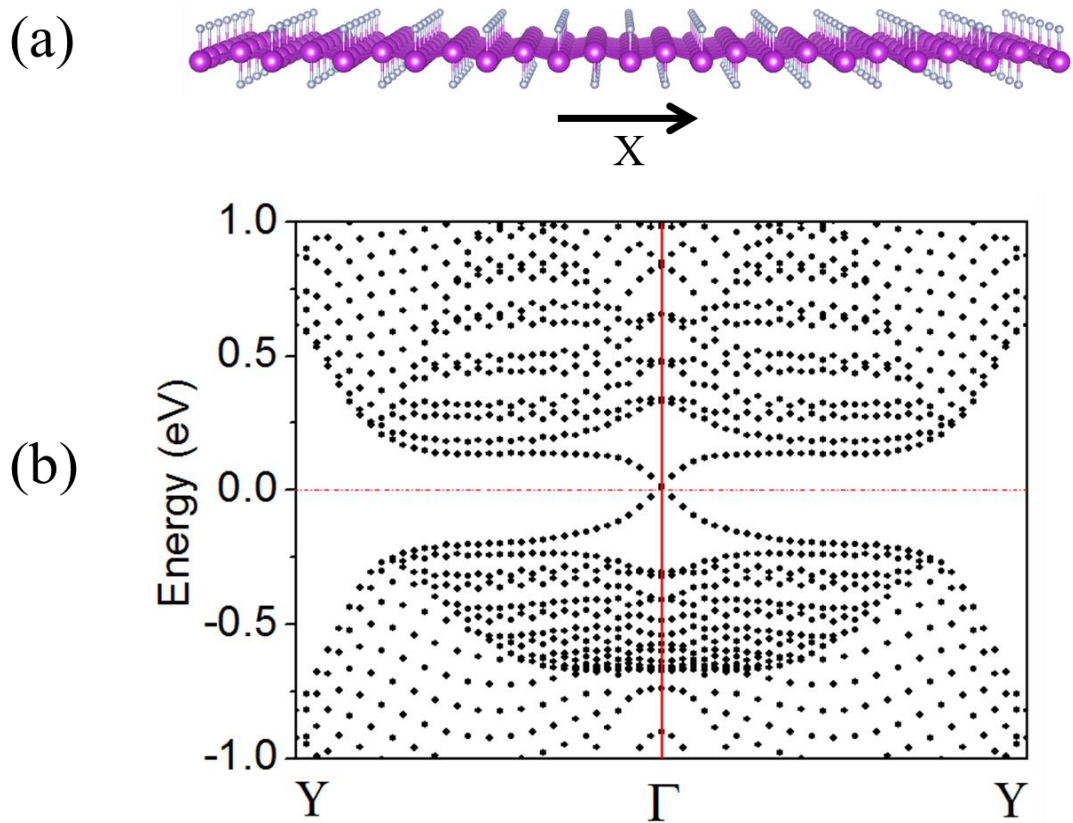

**Supplementary Figure S6. Geometrical and electronic structures of a $Bi_2F_2$ nanoribbon.** (a) Geometry of a nanoribbon constructed along the x axis. The width is 12 times of lattice constant, about 5.4 nm. (b) LDA+SOC band structure of the ribbon.

One can clearly see that two gapless edge states appear in bulk band gap, which cross each other linearly at Γ point. These edge states are mainly contributed by the edge Bi atoms, which further prove that $Bi_2F_2$ is a TI.

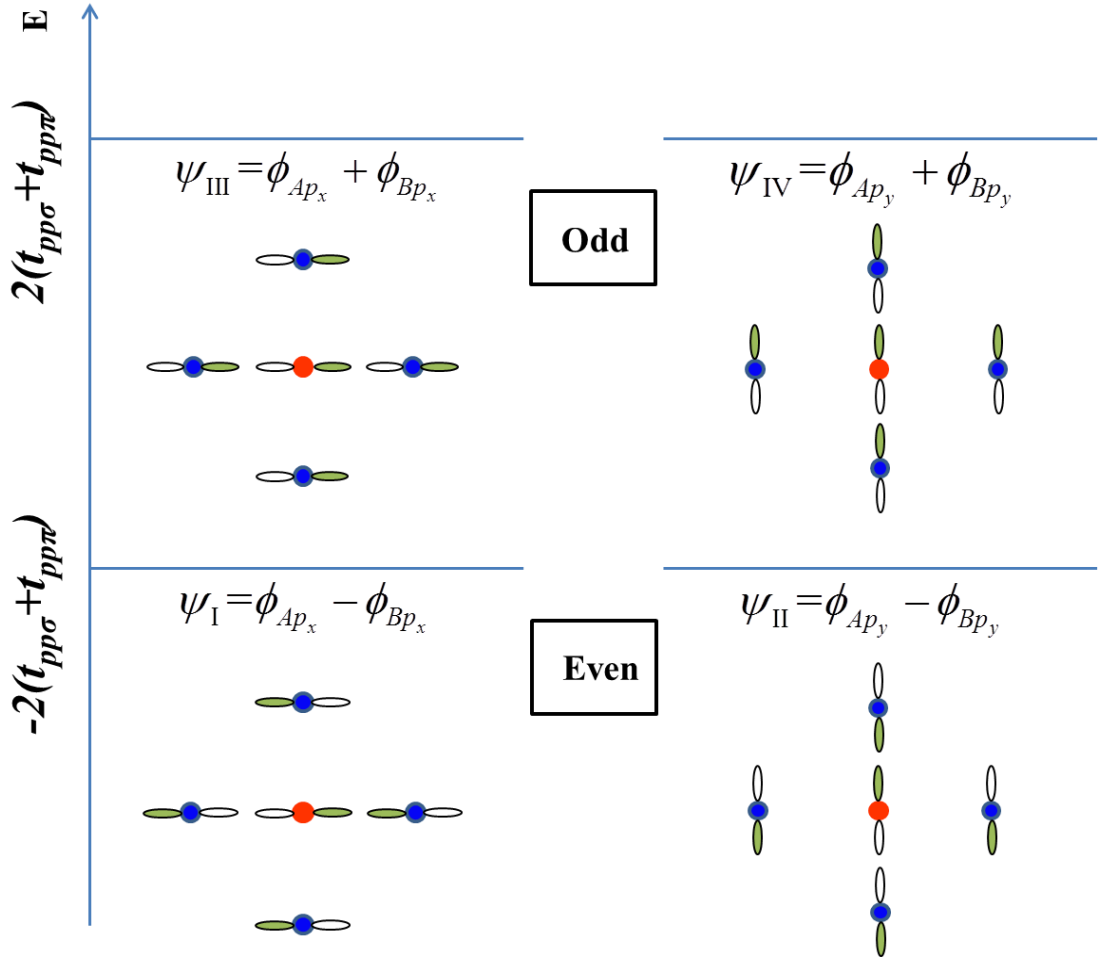

**Supplementary Figure S7. Schematic illustration wave functions at Γ for the square lattices.** Here, only the usual hopping term is considered. The eigenvalues can be expressed as $E_I = E_{II} = -2(t_{pp\sigma} + t_{pp\pi})$ and $E_{III} = E_{IV} = 2(t_{pp\sigma} + t_{pp\pi})$. States I and II with even parity have lower energy than states III and IV with odd parity since $\sigma$ interaction is stronger than $\pi$ interaction. This is the key to the TI behavior in the buckled square lattices.